# Determining Context Factors for Hybrid Development Methods with Trained Models

Jil Klünder[1], Dzejlana Karajic[2], Paolo Tell[3], Oliver Karras[4], Christian Münkel[5], Jürgen Münch[6], Stephen G. MacDonell[7], Regina Hebig[8], and Marco Kuhrmann[9]

[1,4&5]*Leibniz University Hannover,* [2&9]*University of Passau,* [3]*IT University Copenhagen,* [6]*Reutlingen University,*
[7]*Auckland University of Technology,* [8]*Chalmers | University of Gothenburg*

[1]jil.kluender@inf.uni-hannover.de, [2]dzejlana.karajic@gmail.com, [3]pate@itu.dk, [4]oliver.karras@inf.uni-hannover.de,
[5]christian@muenkel.cc, [6]j.muench@computer.org, [7]stephen.macdonell@aut.ac.nz, [8]regina.hebig@cse.gu.se, [9]kuhrmann@acm.org

**Abstract**

*Selecting a suitable development method for a specific project context is one of the most challenging activities in process design. Every project is unique and, thus, many context factors have to be considered. Recent research took some initial steps towards statistically constructing hybrid development methods, yet, paid little attention to the peculiarities of context factors influencing method and practice selection. In this paper, we utilize exploratory factor analysis and logistic regression analysis to learn such context factors and to identify methods that are correlated with these factors. Our analysis is based on 829 data points from the HELENA dataset. We provide five base clusters of methods consisting of up to 10 methods that lay the foundation for devising hybrid development methods. The analysis of the five clusters using trained models reveals only a few context factors, e.g., project/product size and target application domain, that seem to significantly influence the selection of methods. An extended descriptive analysis of these practices in the context of the identified method clusters also suggests a consolidation of the relevant practice sets used in specific project contexts.*

## 1. INTRODUCTION

Determining and accounting for the context in which a development method must be used is among the most challenging activities in process design [1, 4, 23]. For every project, many context factors have to be considered—and the number of such context factors is huge. For instance, Clarke and O'Connor [7] identify 44 major situational factors (with in total 170 sub factors) in eight groups. Kalus and Kuhrmann [17] name 49 tailoring criteria. In this regard, situational factors and tailoring criteria both represent context factors. In both studies, authors do not claim to have explored all factors and discuss that further domain-specific aspects could extend the set of factors identified. Also, in both studies, authors point to issues regarding the mapping of context factors with specific methods and development activities in projects. Such a mapping is usually performed during project-specific process tailoring, which however still seems to be implemented in a demand-driven and experience- based way [21] rather than in an evidence-based manner.

Recent research provides initial evidence on the systematic use of hybrid methods in industry, i.e., methods that are combinations of multiple development methods and practices [19]. In [32], we proposed a statistical construction procedure for hybrid methods, which is grounded in evidence obtained in a large-scale survey among practitioners [22]. Yet, our approach left out context factors and employs usage frequencies to compute base methods and method combinations that build the framework for plugging in sets of development practices. In 2017, we used statistical clustering methods to identify related methods and practices [20]. However, the direct influence of context factors and the influence of latent factors was not included in these previously conducted studies.

**Problem Statement.** Even though available research agrees on the importance of context factors in the construction of development methods for a specific project context, an evidence-based method that helps define the "best-fitting" method for a specific context is missing. This adds a risk to software projects, since inappropriate hybrid methods can affect several risk-dimensions, e.g., unnecessary work, misunderstandings, and "faked" processes [29].

 **Objective.** We aim to understand the role of context factors, to identify the influential among the context factors, and to understand how to integrate such factors in the systematic and evidence-based construction of hybrid development methods. Hence, the overall objective of this paper is to understand which context factors are important when devising hybrid development methods.

**Contribution.** We contribute a study on the role of context factors in the selection of development methods. Using supervised learning, we analyze a large dataset and derive context factors that influence the selection of development methods. In contrast to our previous study [20], we use Exploratory Factor Analysis and Logistical Regression Analysis methods to learn the context factors from data. The study at hand shows that just a few factors seem to have a significant influence on the method clusters, i.e., project/product size, target application domain, and certain criticality factors. The study at hand also provides a novel approach to refine the construction procedure introduced in [32] in which base methods and method combinations have been identified based on their intentional use. The trained models developed in this study provide a new instrument to refine the method presented in [32] by improving the method-cluster construction through learned context factors.

**Outline.** This paper is organized as follows: Section 2 provides an overview of related work. Section 3 presents the research design, before we present our findings in Section 4. In Section 5 we discuss our findings, before we conclude the paper in Section 6.

## 2. RELATED WORK

Determining and balancing relevant context factors is key [1, 4, 23], however, linking context factors with decisions taken by project managers in process selection with the impact of a specific method is not yet well-understood [17]. Software process improvement (SPI) models, like CMMI [8] or ISO/IEC 15504 [12] have sought to establish such links. However, as recent research [21] shows, software-producing organizations and project teams tend to implement SPI as a project-integrated activity rather than implementing it as a planned project-spanning activity and, therefore, explicit considerations whether the applied development method properly addresses the project context step into the background. Since situation-specific process selection is bound to a particular project, company-wide learning is limited and, thus, the risk increases to use an inadequate development method.

In [32], we could show that there are hundreds of process variants, and Noll and Beecham [27] stated that companies often use hybrid methods, but, tend to stay in a specific process category. As there is no "Silver Bullet" [3, 5, 24, 26, 33] and as companies go for highly individualized development methods [15, 32, 34, 35], notably, for becoming more agile, the need for answering the question of which is the best-fitting development method becomes increasingly relevant. However, to answer this question, a deep understanding of context factors and how these drive the selection of development methods is necessary. The paper at hand aims to close this gap by utilizing trained models of context factors. Utilizing the HELENA data [22], we implement a supervised learning strategy that helps predict and recommend a hybrid development method based on the project context.

## 3. RESEARCH DESIGN

We present the research design including a discussion of the threats to validity. Figure 1 provides an overview of the overall research method, which we explain in detail in subsequent sections.

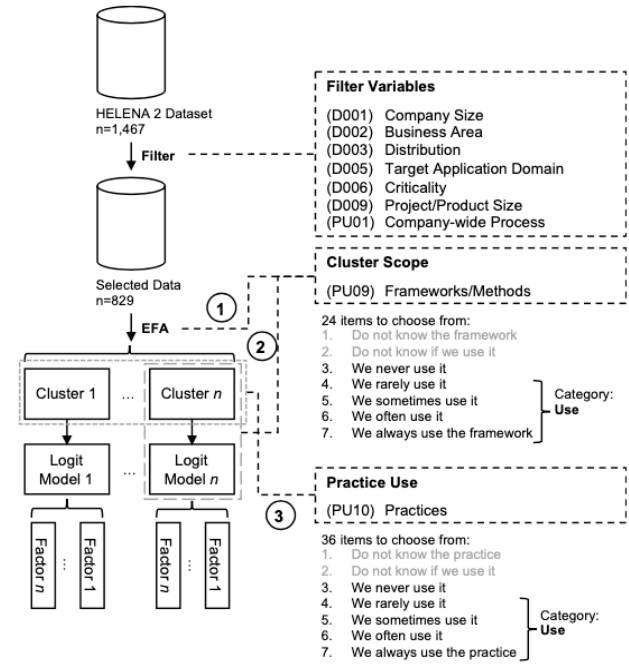

**Figure 1:** Overview of the research method including the data (variables) selected for the study

### 3.1 Research Objective and Research Questions

Our overall objective is to understand which context factors are important when devising hybrid development methods. In particular, we analyze which development methods are related to similar context factors, i.e., we group the set of development methods according to context factors that are related to methods in the set. For this, we pose the research questions presented in Table 1.

### 3.2 Data Collection Procedures

This study uses the HELENA 2 dataset [22] and no extra data was collected. The HELENA 2 data was collected in a large international online survey as described in [19, 32]. Starting in 2015, in three stages, the HELENA survey instrument was incrementally developed and tested. In total, the questionnaire consisted of five parts and up to 38 questions, depending on previously given answers. Data was collected from May to November 2017 following a convenience sampling strategy [31].

**3.2.1 Variable Selection.** In this paper, we focus on the context factors of the development process and their relation to the chosen development methods. Therefore, we only consider answers to selected questions on context factors. This selection defines the base dataset for our analysis. It consists of the parameters D001, D002, D003, D005, D006, D009, and PU01 as shown in Figure 1. All these variables/questions are used in the dataset to describe context-related properties of the actual method use. Furthermore, we excluded all questions that are based on perceptions, such as the degree of agility per project category, or that are defined by the participants, such as the participant's role or the years of experience. Due to an unequal distribution of data points per country [19], we also excluded the country as a context factor from the analysis.

**Table 1:** Overview of the research questions of this study

| | Research Question and Rationale |
|---|---|
| RQ1 | *Which development methods have similar usage contexts?* <br> In the first step, we study clusters of development methods. In contrast to our previously published study [20], we use an *Exploratory Factor Analysis* to build clusters of methods. |
| RQ2 | *Which context factors influence the likelihood of using development methods from a specific set?* <br> In the second step, we study context factors, which are the building blocks for the clusters identified in *RQ1*, in more detail. We build a *Logistic Model* uncovering the factors that influence the likelihood of ending in the respective method clusters. |
| RQ3 | *Which practices are commonly used to extend the method clusters and hence should be taken into consideration when forming a hybrid development method?* <br> After having figured out the clusters of methods in *RQ1*, these methods need to be extended with practices [28, 32], which are the building blocks of development methods. In this study, we are primarily interested in identifying candidate practices and, therefore, we analyze the sets of practices descriptively only. |

**3.2.2 Data Selection.** The survey yielded 1,467 responses, and 691 participants completed the questionnaire. In this study, we use the complete dataset, which includes only answers of the participants that completed the survey according to the selection of variables. This leads to a base population of $n = 836$.

**3.2.3 Data Cleaning.** To prepare the data analysis, we inspected the data and cleaned the dataset. Specifically, we analyzed the data for NA and -9 values indicating that the participants did not provide answers. That is, participants either skipped a question or did not provide an answer to an optional question. Data points containing such values have been analyzed and omitted if the remaining in- formation was insufficient to be included in the statistical analyses. We removed a data point as soon as it contained NA or -9 for at least one variable—including PU09—under investigation. This leads to a final base population of $n = 829$.

## 3.3 Data Analysis Procedures

As illustrated in Figure 1, we implemented a three-staged data analysis procedure, which consists of an Exploratory Factor Analysis as the first step, the construction of a set of Logistic Models in the second step, and a descriptive analysis of practice use in the third step. In subsequent sections, we provide detailed information on the chosen methods and their application.

**3.3.1 Exploratory Factor Analysis.** To answer the first research question (Table 1), we performed an Exploratory Factor Analysis (EFA). An EFA is "a multivariate statistical method designed to facilitate the postulation of latent variables that are thought to underlie—and give rise to—patterns of correlations in new domains of manifest variables" [10]. In the first step of our study, we used an EFA to uncover latent variables or hypothetical constructs. A latent variable cannot be directly observed, but, it can emerge[1] from a set of other observed variables. Specifically, we aim to create clusters of methods, based on the use of the methods concerning a similar degree and similar method combinations. For this, we use the 829 data points, each containing information about the use of methods. We implemented the EFA in the following steps:

**Step 1: Applicability of the EFA.** The first step is to ensure that an EFA is applicable to our dataset. In this regard, the first important criterion for applying an EFA is to answer the question if the correlation matrix of the variables under consideration is the identity matrix for which Bartlett's test [2] is used. If this is the case, the EFA should not be applied. Therefore, we performed Bartlett's test to ensure that an EFA can be applied. Furthermore, we applied the Kaiser-Meyer-Olkin measure (KMO; [16]) to analyze the suitability of our dataset for an EFA. However, as the KMO provides a metric for all potentially relevant variables, we also checked the individual variables using the Measure of Sampling Adequacy (MSA; [16]) for each variable. Finally, we check if the determinant of the correlation matrix is greater than 0.00001 to avoid singularities. In our case, the determinant is 0.00031 > 0.00001.

**Step 2: Calculating the Number of Clusters.** Having ensured that the EFA is applicable to our dataset, in the next step, we calculate the number of clusters (factors) that should be generated by the EFA. A parallel analysis is a common approach for deciding on the number of factors. Often, parallel analysis is combined with the so-called Scree test, which is also known as Cattell's Criterion [6], which, in our case, suggests to build five factors. A double check using the Kaiser criterion [16] resulted in the suggestion to build two factors. To obtain more detailed results, we opted for the Scree test and used the five-factor suggestion to build five method clusters.

**Step 3: Performing the EFA.** Eventually, we performed the EFA using R[2], which constructs the method clusters. We used ordinary, least squares, minres as factoring method with factor loadings ≥ 0.3, since we cannot guarantee normally distributed data. Furthermore, we used Oblimin [13] as rotation method. As an oblique rotation method, Oblimin permits correlations among the constructed sets, and in case of uncorrelated data, rotations produce similar results as orthogonal rotation.

**Quality Evaluation of the Analysis.** To ensure the quality of the results, we calculate the Root Mean Square Error of

---

[1] In this context, a latent variable could be agile, hybrid or traditional, or anything that emerges from clustering the different methods and frameworks.

[2] See: https://www.promptcloud.com/blog/exploratory-factor-analysis-in-r

Approximation (RMSEA), which estimates the discrepancy between the model and the data. Furthermore, we calculate Cronbach's $\alpha$ for each identified cluster to analyze if the clustering of methods is reliable, i.e., if all elements in the cluster calculate the same. For the interpretation of Cronbach's $\alpha$, we use the widely accepted scale: $\alpha \geq 0.9$ is considered excellent (items are highly correlated), $0.9 > \alpha \geq 0.8$: good, $0.8 > \alpha \geq 0.7$: acceptable, $0.7 > \alpha \geq 0.6$: questionable, $0.6 > \alpha \geq 0.5$: poor, and $0.5 > \alpha$ is considered unacceptable.

**3.3.2 Logistic Regression Analysis.** To answer the second research question (Section 3.1), we analyze which context factors influence the likelihood of being allocated to one of the identified clusters as a starting point for deciding on an appropriate development method using the results of [32]. Since this allocation is a binary decision, we use the *Binary Logistic Regression* analysis—a so-called Binary Logistic Model—to calculate the influence of each context factor for the allocation. In the following, we describe the five steps performed in our logistic regression analysis.

**Step 1: Checking Assumptions.** Before applying a logistic regression analysis, several assumptions need to be checked. For our analysis, we check the following four assumptions according to Karras et al. [18]:

1. *The dependent variable is dichotomous.* To fulfil this assumption, a binary outcome needs to be defined. That is, it must be clearly defined whether or not a data point is in a method cluster.

2. *The independent variables are metric or categorical.* All variables used in the model are categorical (Section 3.2.1). All variables are assessed on Likert scales (ordinal answers) or as single-choice options (nominal answers).

3. *In the case of two or more metric independent variables, no multicollinearity is allowed to be present.* This assumption is fulfilled, since we do not have any metric independent variables.

4. *Both groups of the dichotomous dependent variable contain at least 25 elements.* This assumption has to be checked after having defined the binary outcome (see the next Step 2).

**Step 2: Defining the Binary Outcome.** Since we used the raw data as a training set for the Logistic Model, we first define a threshold for accepting a data point for a specific cluster. Note: for each cluster, we consider only those data points that provide information about the use of every method in the cluster. This means that we exclude data points that report for one or more of the cluster's methods that the method is not known or that it is now known if the method is used (see classification in Figure 1). Hence, the investigation of the clusters considers a different subset $n$total of the overall set of data points. To accept a data point for a cluster, we define the criterion "use" through PU09$i \geq 4$ (Figure 1), i.e., the method $i$ was at least rated "rarely used" by the participant [19]. Due to the varying number of methods used by the participants, we defined a relative threshold $> 0.5$, i.e., a data point is added to a cluster if at least 50% of the cluster's methods are used in the data point. The resulting number of data points in a cluster is called $n$using cluster in the following (see Table 3 for an overview).

**Step 3: Data Preparation for the Logistic Analysis.** For each cluster identified in the EFA (Section 3.3.1), we built one Logistic Model. Each Logistic Model aims at identifying variables that influence the likelihood of finding suitable methods in the associated cluster of methods (Figure 1). To build the Logistic Models, data needs to be prepared. For each cluster identified, we therefore independently performed the following steps:

1. We removed all data points that did not match the "use" criterion defined in Step 2, i.e., we removed all data points with PU09$i < 3$ (don't know the method $i$ and don't know if we use it). The whole data point was removed at the first occurrence of a rating $< 3$, as we cannot draw conclusions on the method use. This rigorous decision helps reduce noise in the data as we only include complete data points in the analysis.

2. As the logistic analysis requires factorized variables, i.e., every possible answer option of the considered variables is treated as a single categorical variable. Figure 2 illustrates the factorization for the variable D001. This procedure was applied to the variables D001, D003, D009, and PU01. The remaining variables described in Section 3.2.1 are already presented and interpreted as categorical variables.

3. Some multiple-choice questions (e.g., D002, D005, and D006; Figure 2, [22]) had an option "other" to provide extra information through free-text answers. Due to the diversity and low number of reoccurring answers to these options, we decided to exclude these answers from the analysis.

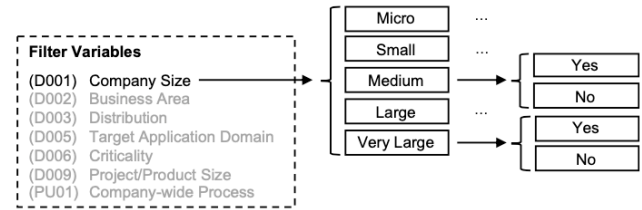

**Figure 2:** Variable factorization to prepare the Logit analysis

**Step 4:** Build the Models. We build all binary logistic models with R. All models considered in total 48 variables (constructed as described in Step 3). Further details for the significant predictors of the models can be found in Section 4.2.

**Step 5:** Evaluate the Model. To ensure interpretability of the results, we performed multiple steps for the evaluation as suggested by Peng et al. [30]:

1. We evaluated the overall model quality using the Likelihood ratio test, which compares the model's results with the results given by the (intercept-only) null model. Since the null hypothesis states that there is no difference between the logistic regression model and the null model, this test needs to be significant.

2. We performed Wald's test [11] to analyze the statistical significance of the individual predictors.

This step is necessary for the interpretation of the model's results.

3. We calculated goodness-of-fit statistics using the Hosmer- Lemeshow test (HL) and Nagelkerke's $R2$ [25]. These statistics analyze the fit of the logistic model against the actual out- comes, i.e., the statistics indicate if the model fits the original data. The HL-test must not be significant as it tests the null hypothesis that the model fits the data. Nagelkerke's $R2$ calculates how much variability in the dataset can be explained by the logistic model—the closer $R2$ is to 1, the more variability can be explained by the model. This value can be converted into Cohen's effect size $f$ [9] to assess the practical relevance of the results.

4. We validated the predicted probabilities to calculate the ac- curacy of the Logistic Model. Accuracy can be expressed as a measure of association ($c$-statistic) and a classification table that summarizes true and false positives/negatives. The $c$- statistic is a measure $0.5 \leq c \leq 1$, with 0.5 meaning that the model is not better than a random prediction and 1 meaning that all pairwise assignments of elements in and not in the cluster are always correct. A confusion matrix summarizes the results of applying the Logistic Model to the dataset. For each entry in the dataset, the model calculates the predicted probabilty and classifies the entry into one of the two possible groups of the dependent variable. Afterwards, it is possible to calculate the sensitivity and the specificity of the model.

**3.3.3 Descriptive Analysis on Practice Use.** To answer the third research question, we analyzed the sets of practices (Figure 1, PU10) for all data points that are defined to contribute to a respective method cluster (Section 3.3.2–Step 2). For this, we calculate the share of practices per cluster. Again, we only consider data points matching the "use" criterion $PU10i \geq 4$ (see Section 3.3.2–Step 2 and Figure 1) and we use a 85% threshold [32] to consider a practice commonly used within a cluster of methods.

**3.4 Validity Procedures and Threats to Validity**

The research presented in this paper is subject to some limitations and threats to validity, which we discuss using the classification by Wohlin et al. [36].

**3.4.1 Construct Validity.** Given that the dataset used in this analysis emerged from an online survey, we had to deal with the risk of misunderstood questions leading to incomplete or wrong answers. To mitigate this risk and as described in Section 3.2, the questionnaire was iteratively developed, including translations into three languages by native speakers [19]. Several methods are related to one another and were partially built on top of each other. Thus, participants, e.g., using Scrum, are likely to identify their development process to be Iterative as well, which could have resulted in false positives during the identification of the hybrid methods. Another risk emerges from the chosen convenience sampling strategy [31] to distribute the questionnaire, which potentially introduced errors due to participants not reflecting the target population. Given the meaningful results of the analysis of free-text answers [19], we are confident that this threat can be considered mitigated.

**3.4.2 Internal Validity.** The selection of variables, which emerged from the limitation to context factors, and the cleaning of the data as described in Section 3.2 can influence the results. The variables, in conjunction with the methods as variables under investigation, defined the basic dataset. Based on this dataset, we removed all incomplete data points. A data point was considered as incomplete as soon as one of the respective questions was not answered. This reduced the overall sample size, but, we considered the flawed interpretations due to missing answers more severe. We followed the same approach for the definition of data points for the second step of the analysis. That is, we also decided conservatively on the inclusion of data points and removed all data points that did not match the "use" criterion defined in Section 3.3.2–Step 2, which reduced the sample sizes, but, we decreased the risk of flawed interpretations. Finally, all steps of the data analysis were performed by two researchers, and two more researchers not involved in the data analysis thoroughly reviewed each step. Therefore, we are confident that the analyses are well documented (for replication) and robust.

**3.4.3 Conclusion Validity.** The interpretation of the statistical tests is based on a significance level of $p \leq 0.05$. Nevertheless, before interpreting the results, we included several reliability checks in the analysis, including the calculation of Cronbach's $\alpha$ for internal reliability of the found sets, measures for error rates (RMSEA), as well as the thorough evaluation of the logistic models as described in Section 3.3.2–Step 5. We used a 85% threshold for the extension of the method sets with practices. This threshold was defined in [32] on the same dataset. Changing this threshold would impact the results and limit the usability of [32] as a baseline. Further research is thus necessary to increase the results' reliability.

**3.4.4 External Validity.** Our results emerge from a large-scale study representing development methods of a large number of companies with different context factors and in different environments. Yet, we cannot guarantee that our results are correct and applicable for each company. Nevertheless, we found evidence that some context factors tend to be more important than others. These may be taken into account when defining hybrid development methods.

# 4. Results

We present the results of our study following the three steps of the data analysis shown in Figure 1 and described in Section 3.3.

**4.1 Exploratory Factor Analysis**

As described in Section 3.3.1, the Exploratory Factor Analysis was performed in different steps for which we present the results in this section.

**4.1.1 Step 1: Applicability of the EFA.** Before performing the EFA, we ensured that it is applicable to our dataset. For this, we calculated the MSA for each variable. These checks resulted in an overall MSA = 0.9, which is considered good to very good. Only the two values for "Scrum" and "Kanban" resulted in medium suitability. Nevertheless, the smallest result in the dataset was MSA = 0.78 (medium) and, therefore, EFA can be applied to our dataset.

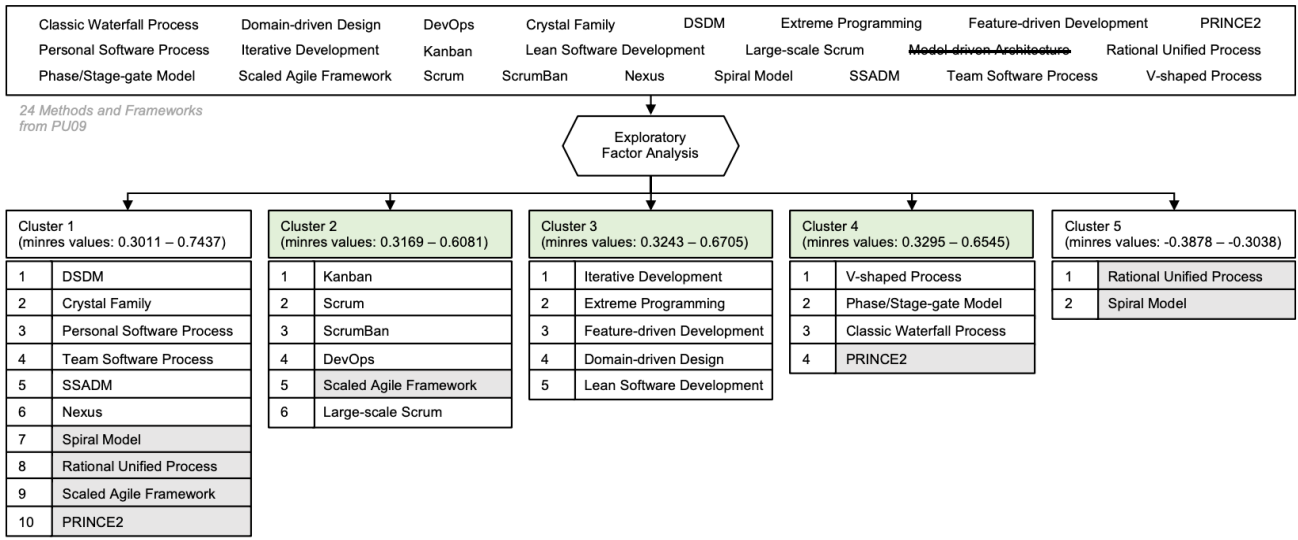

**Figure 3:** The five resulting clusters of the Exploratory Factor Analysis (including value ranges of the minres-algorithm, grey cells highlight methods that are relevant in multiple clusters)

**4.1.2 Step 2: Calculating the Number of Clusters.** To calculate the number of factors to be considered, we performed a parallel analysis. The Scree test suggests five factors. We followed this suggestion to obtain more fine-grained results and constructed five method clusters.

**4.1.3 Step 3: Performing the EFA.** Performing the EFA resulted in the method clusters shown in Figure 3. We removed all elements with loadings $|l| < 0.3$, which is a common practice, e.g., "Model-driven Architecture". A greater loading-value represents a more important position in the set, which is illustrated in Figure 3 by the "rank" of the methods. For example, "Kanban" was "more important" to define Cluster 2 than "Large-scale Scrum". The factor loading indicates the strength and direction of a factor on a measured variable and, therefore, these values are particularly important in the interpretation of the sets.

**4.1.4 Quality Evaluation of the Analysis.** To check the quality of the model, we calculated the RMSEA fit index as 0.058. A value between 0.05 and 0.08 constitutes a good fit of the model. Furthermore, Table 2 summarizes the results of the Cronbach's $\alpha$ values for each cluster. While the clusters 1, 2 and 3 are well defined according to the internal reliability, Cluster 4 is questionable, and Cluster 5 is poor. Hence, the clusters 4 and 5 need to be treated with care.

**Table 2:** Cronbach's $\alpha$ reliability check for the clusters

| Cluster | Cronbach's $\alpha$ | Interpretation | Logistic Analysis? |
|---|---|---|---|
| Cluster 1 | 0.86 | good | No |
| Cluster 2 | 0.70 | acceptable | Yes |
| Cluster 3 | 0.73 | acceptable | Yes |
| Cluster 4 | 0.66 | questionable | Yes |
| Cluster 5 | 0.56 | poor | No |

**4.2 Logistic Regression Analysis**

As described in Section 3.3.2, the Logistic Regression Analysis was performed in different steps. In the EFA presented in Section 4.1, we identified five clusters having relationships to similar context factors. In this section, we present the analysis of the correlation of context factors and the clusters.

**4.2.1 Data Preparation for the Logistic Analysis.** We prepared the dataset for each cluster identified in the EFA following the steps described in Section 3.3.2–Step 3. The data preparation yielded the distribution of datasets shown in Table 3: column $n$total shows the number of data points that provide information about use or missing use for all methods in the cluster. The column $n$using cluster shows the number of data points that report using more than half of the clusters' methods (see Section 3.3.2). Finally, column $n$rest includes the remaining data points of $n$total that are not part of $n$using cluster. As we had less than 25 data points for Cluster 1, we could not apply the logistic regression analysis to this cluster.

**Table 3:** Data points per cluster used for the Logistic Models

| Cluster | $n_{total}$ | $n_{using\ cluster}$ | $n_{rest}$ |
|---|---|---|---|
| Cluster 1 | 135 | 16 | 119 |
| Cluster 2 | 281 | 130 | 151 |
| Cluster 3 | 352 | 217 | 135 |
| Cluster 4 | 209 | 53 | 156 |
| Cluster 5 | 376 | 65 | 311 |

**4.2.2 Results.** Based on the clusters identified in the EFA (Figure 3) and the results of the data preparation shown in Table 3, we performed the logistic regression analysis using R. In the following, we present the results of these analyses per cluster.

**Cluster 1.** For the small number of 16 data points (Table 3) using more than 50% of the methods in Cluster 1, one condition to apply the logistic regression as described in Section 3.3.2 was violated. Therefore, it was not possible to build a Logistic Model for Cluster 1.

**Cluster 2.** The first step was to calculate the factorized variables as described in Figure 2. The significant results (Wald's tests) are summarized in Table 4. The table shows that none of the factorized variables significantly influences the likelihood of using more than 50% of the methods in Cluster 2. The other results are further refined in Table 5, which summarizes the significant results of the Logistic Model.

**Table 4:** Wald's test results of the Logistic Model for Cluster 2

| **Var.** | | $\chi^2$ | df | $P(>\chi^2)$ | **Interpretation** |
|---|---|---|---|---|---|
| D001 | Company Size | 4.3 | 4 | 0.37 | not significant |
| D003 | Distribution | 7.0 | 3 | 0.07 | not significant |
| D009 | Project/Prod. Size | 1.6 | 4 | 0.8 | not significant |
| PU01 | Comp.-w. Process | 0.054 | 2 | 0.97 | not significant |

**Table 5:** Significant results of the Logistic Model for Cluster 2

| **Var.** | | **Est.** | **Std. Err.** | **z-value** | $P(>|z|)$ |
|---|---|---|---|---|---|
| $D003_3$ | Distr. Continent | 1.02 | 0.48 | 2.105 | 0.0353 |
| $D005_{04}$ | Defense Systems | 3.52 | 1.63 | 2.156 | 0.0311 |
| $D005_{15}$ | Space Systems | -3.85 | 1.89 | -2.035 | 0.0419 |

Based on the results from Table 4 and Table 5, we conclude the following statements: Based on the logistic model, the likelihood that a company uses more than 50% of the methods contained in Cluster 2 is...

1. ... positively related to companies that are distributed across one continent ($D003_3$, $p < 0.05$, Est. = $1.02 > 1$). That is, companies working in this area tend to use more agile development methods, including the agile scaling frameworks, e.g, SAFe or LeSS.

2. ... positively related to companies active in the defense systems domain ($D005_{04}$, $p < 0.05$, Est. = $3.52 > 1$). That is, companies working in this area tend to use more agile methods.

3. ... negatively related to companies active in the space systems domain ($D005_{15}$, $p < 0.05$, Est. = $-3.85 < 1$). That is, companies working in this area tend to avoid using agile methods.

Finally, as described in Section 3.3.2–Step 5, we evaluated the model. The first step was to conduct the Likelihood Ratio Test, which confirmed that there is no significant improvement comparing the built model with the null model ($\chi^2 = 36.902$, $p = 0.06197 < 0.1$). However, at a significance level of $p = 0.1$, there is a significant improvement. Therefore, the results of the model need to be taken with care. In the second step, a z-test was performed to analyze the statistical significance of the individual predictors. Table 4 and Table 5 show three significant variables, i.e., predictors for the likelihood of using methods from Cluster 2. In the analysis of the goodness-of-fit statistics, the Hosmer-Lemeshow test did not support the claim that the model does not fit the data ($\chi^2 = 9.0373$, df = 8, $p = 0.3392 > 0.05$). Nagelkerke's $R2$ resulted in 0.272, i.e., almost 27% of the variability in the dataset can be explained with the logistic model, and the resulting effect size of $f = 0.282$ indicates a medium effect [9].

To validate the predicted probabilities, we defined a threshold using the share of data points contributing to the cluster (Table 3), i.e., $1 - \frac{130}{281} = 0.537$, which is the relative probability of not being in the cluster. Using this relative probability, we computed the confusion matrix with a sensitivity of 79.47% and a specificity of 53.85%. The c-statistic was calculated with 0.7666, which indicates a good result as it states that for approx. 75% of all possible pairs of data points, the model correctly assigned the higher probability to those in the cluster.

**Table 6:** Wald's test results of the Logistic Model for Cluster 3

| **Var.** | | $\chi^2$ | df | $P(>\chi^2)$ | **Interpretation** |
|---|---|---|---|---|---|
| D001 | Company Size | 5.4 | 4 | 0.25 | not significant |
| D003 | Distribution | 3.1 | 3 | 0.38 | not significant |
| D009 | Project/Product Size | 7.6 | 4 | 0.11 | not significant |
| PU01 | Comp.-wide Process | 0.16 | 2 | 0.92 | not significant |

**Table 7:** Significant results of the Logistic Model for Cluster 3

| **Var.** | | **Est.** | **Std. Err.** | **z-value** | $P(>|z|)$ |
|---|---|---|---|---|---|
| $D005_{16}$ | Telecom. | -1.32 | 0.66 | -2.002 | 0.0453 |

**Cluster 3.** The first step was to calculate the factorized variables as described in Figure 2. The results (Wald's tests) are summarized in Table 6. The table shows that none of the factorized variables significantly influences the likelihood of using more than 50% of the methods in Cluster 3. Table 7 summarizes the significant results of the Logistic Model for the refinement of the individual variables.

Based on the results from Table 6 and Table 7, we conclude the following statement: Based on the logistic model, the likelihood that a company uses more than 50% of the methods contained in Cluster 3 is negatively related to companies active in the domain of Telecommunication ($D005_{16}$, $p < 0.05$, Est. = $-1.32 < 1$). That is, companies working in this area tend to avoid using agile methods.

As described in Section 3.3.2–Step 5, we evaluated the model. We conducted the Likelihood Ratio Test, which confirmed that there is a significant improvement comparing the built model with the null model ($\chi^2 = 71.124$, $p = 0.01671 < 0.05$). A z-test was performed to analyze the statistical significance of the individual predictors. Table 6 and Table 7 show one significant variable, i.e., one predictor for the likelihood of (not) using methods from Cluster 3. In the analysis of the goodness-of-fit statistics, the Hosmer-Lemeshow test did not support the claim that the model does not fit the data ($\chi^2 = 6.3067$, df = 8, $p = 0.6129 > 0.05$). Nagelkerke's $R2$ resulted in 0.249, i.e., almost 25% of the variability in the dataset can be explained with the logistic model, and the resulting effect size of $f = 0.257$ indicates a medium effect [9].

To validate the predicted probabilities, we defined a threshold using the share of data points contributing to the cluster (Table 3), i.e., $1 - \frac{217}{352} = 0.384$, which is the relative probability of not being in the cluster. Using this relative probability, we computed the confusion matrix with a sensitivity of 28.89% and a specificity of 94.01%. The c-statistic was calculated with 0.7434, which indicates a good result as it states that for approx. 75% of all possible pairs of data points, the model correctly assigned the higher probability to those in the cluster.

**Cluster 4.** At first the factorized variables were calculated as de- scribed in Figure 2. The significant results (Wald's tests) are shown in Table 8. The table shows that the Project/Product Size (D009) significantly influences the likelihood of using more than 50% of the methods in Cluster 4. This particular influence is refined in Table 9, which summarizes the significant results of the Logistic model.

**Table 8:** Wald's test results of the Logistic Model for Cluster 4

| Var. | | $\chi^2$ | df | $P(>\chi^2)$ | Interpretation |
|---|---|---|---|---|---|
| D001 | Company Size | 5.3 | 4 | 0.26 | not significant |
| D003 | Distribution | 3.6 | 3 | 0.31 | not significant |
| D009 | Project/Product Size | 10.3 | 4 | 0.035 | significant |
| PU01 | Comp.-wide Process | 1.8 | 2 | 0.41 | not significant |

Based on the results from Table 8 and Table 9, we conclude the following statements: Based on the logistic model, the likelihood that a company uses more than 50% of the methods contained in Cluster 4 is. . .

1. . . . negatively related to companies active in the domain of Web Apps. and Services (D00517, $p < 0.05$, Est. $= -1.86 < 1$). That is, companies working in this area tend to avoid using traditional development methods.

2. . . . positively related to the size of a product or project – class: Small (D0092, $p < 0.05$, Est. $= 4.65 > 1$). Running small projects (effort: 2 person weeks–2 person months) increases the likelihood of using methods from Cluster 4, i.e., small projects tend to use traditional development methods.

Finally, as described in Section 3.3.2–Step 5, we evaluated the model. The first step is to conduct the Likelihood Ratio Test, which con- firmed that there is a significant improvement comparing the built model with the null model ($\chi^2 = 73.827, p = 0.00971 < 0.05$). A $z$-test was performed to analyze the statistical significance of the individual predictors. Table 8 and Table 9 show two significant variables, i.e., two predictors for the likelihood of using methods from Cluster 4. In the analysis of the goodness-of-fit statistics, the Hosmer-Lemeshow test did not support the claim that the model does not fit the data ($\chi^2 = 6.2849$, df $= 8$, $p = 0.6154 > 0.05$). Nagelkerke's $R^2$ resulted in 0.439, i.e., almost 44% of the variability in the dataset can be explained with the logistic model, and the resulting effect size of $f = 0.489$ indicates a large effect [9].

**Table 9:** Significant results of the Logistic Model for Cluster 4

| Var. | | Est. | Std. Err. | z-value | $P(>|z|)$ |
|---|---|---|---|---|---|
| D00517 | Web Appl./Svc. | -1.86 | 0.88 | -2.122 | 0.0338 |
| D0092 | Small Product | 4.65 | 2 | 2.324 | 0.0201 |

---

[3] Please note that we did not execute the same construction method based on usage frequencies and agreement levels as implemented in [32]. In this study, we only used the exploratively identified thresholds for the agreement levels. Yet, we did not apply the combined-set construction to identify process

To validate the predicted probabilities, we defined a threshold using the share of data points contributing to the cluster (Table 3), i.e., $1 - \frac{53}{209} = 0.746$, which is the relative probability of not being in the cluster. Using this relative probability, we computed the confusion matrix with a sensitivity of 98,08% and a specificity of 15.09%. The $c$-statistic was calculated with 0.8586, which indicates a good result as it states that for approx. 86% of all possible pairs of data points, the model correctly assigned the higher probability to those in the cluster.

**Cluster 5.** Due to the poor reliability of Cluster 5 (see Table 2), we did not build the Logistic Model for this cluster.

### 4.3 Practice Use

The last step in our analysis (Figure 1) was the analysis of practices used in the method clusters identified in the EFA (Section 4.1). To determine the clusters of practices, we implemented the (descriptive) analysis method as described in Section 3.3.3[3]. Please note that this part of the analysis is an exploratory analysis in which we are primarily interested in learning if there is an effect on the selection of practices in the context of our factor analysis at all, and if we can observe converging subsets of practices that, eventually, can form clusters of core practices as building blocks of hybrid development methods as identified in [32]. Figure 5 visualizes the outcome of the assignment of practices to clusters. Even though the clusters 1 and 5 could not be considered for the logistic analysis (Table 2 and Table 3), we present the sets of practices for all clusters.

#### 4.3.1 Practices for Analyzed Clusters.
Figure 5 highlights the three clusters of methods for which we implemented Logistic Models. For these three clusters, similar to our findings from [32], we see that a maximum of 26 out of 36 practices find an 85% agreement regarding their use in the context of the method clusters. That is, we also observe some "preferences" regarding the use of practices. For instance, in all three clusters, we find the core practices "Code Review", "Coding Standards", and "Release Planning" (highlighted in Figure 5), which were identified as the least common denominator in [32]. As illustrated in Figure 4, these three practices build one core component of constructing hybrid development methods.

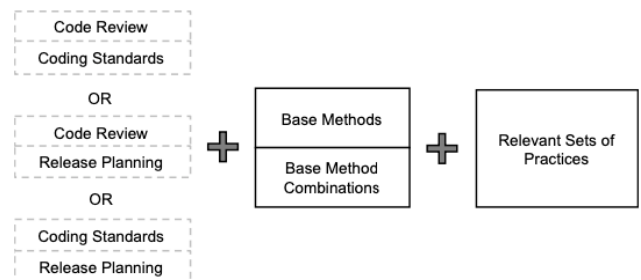

**Figure 4:** Simplified construction procedure [32] with core practices, base methods and combinations, and practice sets for the respective method combinations

---

variants. This requires adjustments of the construction procedure and remains subject to future work (Section 6).

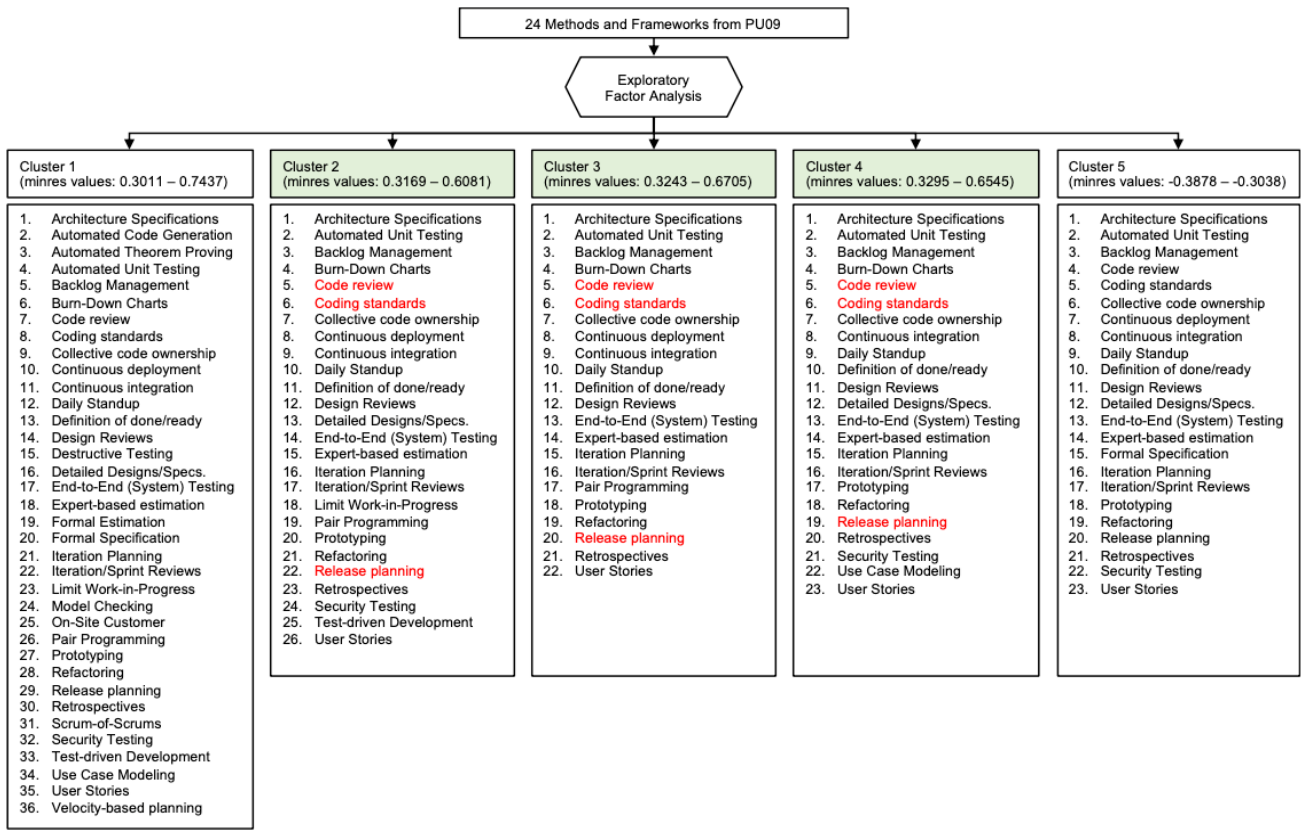

**Figure 5:** Results of the cluster-practice assignment using a 85% agreement level [32] regarding the use of a practice

Together with the initially identified preferences (agreement levels ≥ 0.85) of the practice use in the study at hand, we therefore expect converging sets of practices and combinations of practices for devising hybrid development methods. While the study [32] was limited to only one variable "intentional use of hybrid methods", this study adds further relevant context factors and provides a means for a refined and context-sensitive identification of the base methods and their combinations.

**4.3.2 Practices for Excluded Clusters.** Figure 5 also includes Cluster 1 and Cluster 5, which have been excluded from the logistic regression analysis. However, we can make some interesting observations. First, Cluster 5, which was excluded from the logistic regression analysis due to poor reliability, has a reduced and con- verging set of practices assigned. This set of practices also includes the core practices [32] and, therefore, this cluster remains a candidate for further investigation. The second observation is that Cluster 1 has all practices assigned. As discussed in Section 3.3.2 and Section 4.2.2, the number of elements in Cluster 1 is too small to draw meaningful conclusions. Figure 5 illustrates this by assigning a complete and unfiltered list to the Cluster 1. This finding shows the necessity for further research to grow and improve the data basis.

## 5. DISCUSSION

From the predefined list of 24 methods, we extracted five clusters with methods that are correlated with similar context factors. These clusters consist of mostly agile methods (*Cluster 2*), mostly or completely traditional methods (*Cluster 4* and *Cluster 5*) or both (*Cluster 1* and *Cluster 3*). The clusters formed the basis for a logistic regression analysis to study context factors that influence (i.e., increase or decrease) the likelihood of using more than 50% of methods belonging to the respective cluster. These analyses revealed few significant influence factors: *distributed development on one continent*, target application domains *defense systems*, *space systems*, *telecommunications* and *web applications and services*, and *project/product size: small*. However, we found few contextual factors only that support conclusions or at least assumptions on the used development methods. Therefore, we argue that there must be further factors influencing the choice of development methods, which could explain why the definition of a suitable development method is that complicated. As shown in [19], most development methods emerge from experience. However, experience can only take effect when having "something" in place that can be adjusted based on experience, whereas starting from scratch is difficult.

The results of our study provide support for devising hybrid development methods by identifying factors that influence the choice of development methods. However, our results should not be over-interpreted. They represent an initial guideline for defining a hybrid development method. For instance, our findings help find a starting point for selecting base methods or method combinations to define a hybrid development method. Nevertheless, compared with [7, 17] (44 and 49 factors), we could only identify a small number of context factors. Future research is thus strongly required, notably, to study the remaining known factors for which we—so far—could not draw any conclusion. A deeper knowledge about the context factors will also lay the foundation for developing improved tailoring instruments that help project managers define suitable project-specific development methods using a systematic approach in combination with experience and continuous learning [21].

A second key finding is that we can support the claim that practices are the real building blocks of development approaches [14, 32]. For the three analyzed clusters, we could identify at least 22 practices with an agreement level $\geq 85\%$. This indicates that practices might be context-dependent, which implies that focusing on methods only is insufficient. Further research is necessary to gain deeper insights on the role of practices in method development.

## 6. CONCLUSION

In this paper, we studied the use of hybrid methods based on 829 data points from a large-scale international survey. Using an Exploratory Factor Analysis, we identified five clusters of methods. We used these clusters as dependent variables for a Logistic Regression Analysis to identify contextual factors that are correlated with the use of methods from these clusters. The analysis using trained models reveals that only a few factors, e.g., project/product size, and target application domain, seem to significantly influence the method selection. An extended descriptive analysis of the practices used in the identified method clusters also suggests a consolidation of the relevant practice sets used in specific project contexts.

Our findings contribute to the evidence-based construction of hybrid methods. As described in Section 4.3.1, our results provide a means to learn relevant context factors, which can be used to derive base methods and method combinations that, themselves, are a core component of a construction procedure for hybrid methods [32]. That is, a hybrid development methods can be constructed using a set of context factors going beyond the so far used frequency-based construction procedure. Furthermore, an improved knowledge of such context factors will also contribute to better understand and define powerful tailoring mechanisms that help define development methods for specific project situations to reduce overhead introduced through inadequate project-specific processes.

## ACKNOWLEDGMENTS

We thank all the study participants and the researchers involved in the HELENA project for their great effort in collecting data.